\documentclass[showpacs,showkeys,aps,floatfix]{revtex4}

\usepackage{amsmath}    
\usepackage{color}      
\usepackage{subfigure}  
\usepackage{hyperref}   
\usepackage{epsfig} \usepackage{graphicx} \usepackage{color}
\usepackage{amsmath, amsthm}

\def\ARAA{{Ann. Rev. Astron. \& Astrophys.}}
\def\ApJ{{Astrophys. J.}}
\def\ApJL{{Astrophys. J. Lett.}}

\def\ApP{{Astropart. Phys.}}
\def\AA{{Astron. \& Astroph.}}

\def\JCAP{{Journ. of Cosmol. \& Astropart. Phys.}}

\def\MNRAS{{Month. Not. Roy. Astr. Soc.}}
\def\Nature{{Nature}}

\def\PLB{{Phys. Lett. B}}
\def\PR{{Phys. Rev.}}

\def\PRL{{Phys. Rev. Lett.}}
\def\RMP{{Rev. Mod. Phys.}}

\begin{document}

\title{Centaurus A: the one extragalactic source of cosmic rays with
  energies above the knee}

\author {Peter L. Biermann} \affiliation{Max-Planck-Institute for
  Radioastronomy, Auf dem H{\"u}gel 69, 53121 Bonn, Germany}
\altaffiliation[Also at ]{Physics Dept., K.I.T. \  Karlsruhe, Germany}
\altaffiliation{Dept.\ Physics \& Astronomy, Univ.\ Alabama,
  Tuscaloosa, AL, USA} \altaffiliation{Dept. Physics, Univ.\ Alabama at
  Huntsville, AL, USA} \altaffiliation{Dept.\ Physics \& Astronomy,
  Univ.\ Bonn, Germany} \author {Vitor de Souza} \affiliation
               {Universidade de S$\tilde{a}$o Paulo, Instituto de
                 F\'{\i}sica de S$\tilde{a}$o Carlos, Brazil}
               \date{\today}

\begin{abstract}

 The origin of cosmic rays at all energies is still uncertain. In this
 paper we present and explore an astrophysical scenario to produce
 cosmic rays with energy ranging from below $10^{15}$ to $3 \times
 10^{20}$ eV. We show here that just our Galaxy and the radio galaxy
 Cen A, each with their own galactic cosmic ray particles, but with
 those from the radio galaxy pushed up in energy by a relativistic
 shock in the jet emanating from the active black hole, are sufficient
 to describe the most recent data in the energy range PeV to near ZeV.
 Data are available over this entire energy range from the experiments
 KASCADE, KASCADE-Grande and Pierre Auger Observatory. The energy
 spectrum calculated here correctly reproduces the measured spectrum
 beyond the knee, and  contrary to  widely held expectations, no other
 extragalactic source population is required to explain the data, even
 at energies far below the general cutoff expected at $6 \times
 10^{19}$ eV, the Greisen-Zatsepin-Kuzmin turn-off due to interaction
 with the cosmological microwave background.  We present several
 predictions for the source population, the cosmic ray composition and
 the propagation to Earth which can be tested in the near future.

\end{abstract}

\keywords{cosmic rays; radio galaxies; interstellar matter }

\maketitle

\section{Introduction}

Cosmic rays have been originally discovered in 1912/13 by Hess (Hess
1912) and Kohlh{\"o}rster  (Kohlh{\"o}rster 1913) and still today we
have no certainty where they come from. Their overall spectrum has
been shown to be essentially a power-law with a bend down near
$10^{15}$ eV, called the knee, and a turn towards a new flatter
component near $\sim \, 3 \times 10^{18}$ eV, called the ankle, with a
final turn-off just around $10^{20}$ eV, summarized in (Gaisser \&
Stanev 2008).

It is thought, that the component below about $3 \times 10^{18}$ eV is
Galactic and the component above this energy is extragalactic, on the
basis that particles above such an energy would be hard to contain and
isotropize in the magnetic fields in the interstellar medium disk of
the Galaxy.  Different astrophysical scenarios have been proposed to
explain the Galactic and the extragalactic components of the cosmic
radiation, see the overview (Stanev 2010a, 2010b).

The basic paradigm for Galactic cosmic rays has been acceleration in
the shock waves caused by supernova explosions (Baade \& Zwicky 1934).
The process of acceleration is diffusive shock acceleration (Fermi
1949) and it is based on the compression experienced by particles that
get reflected by magnetic irregularities from both sides of a shock in
an ionized magnetic plasma (Drury 1983).  Supernovae are exploding
stars and they may explode either directly into the interstellar
medium or into the stellar wind of the predecessor star (Woosley
2002).  Lagage \& Cesarsky (Lagage \& Cesarsky 1983) showed that
acceleration at the shocks caused explosions into the normal
interstellar medium cannot reach even the energies at the knee.  Heavy
nuclei can be accelerated up to about $10^{18}$ eV  in Galactic
sources, such as supernova explosions of massive stars which explode
into their wind (V\"olk \& Biermann 1988), OB-star super-bubbles,
gamma ray bursts (Dermer 2004) or micro-quasars, active accreting
black holes in stellar binary systems.

For higher energies ($> 10^{18}$ eV) extragalactic sources are the
most accepted candidates. Nearby active radio galaxies with black hole
activity were first proposed by Ginzburg \& Syrovatskii (Ginzburg \&
Syrovatskii 1963) (see also, e.g., (Lovelace 1976) and (Biermann \&
Strittmatter 1987)) as possible sources. Gamma ray bursts in other
galaxies have also been suggested by Waxman (Waxman 1995) and Vietri
(Vietri 1995). The radio galaxy Cen A is a prime example of a possible
astrophysical source (Anchordoqui et al. 2011).  The interaction of
high energy particles with the microwave background limits the
distance of the sources (Greisen 1966; Zatsepin \& Kuzmin 1966; Allard
et al. 2008; Stanev 2010b).  For protons the energy at which the all
particle spectrum from many sources should turn off is estimated to be
$6 \times 10^{19}$ eV and for other chemical elements the energy turn
off is lower (Allard et al. 2008). As the interaction distance becomes
very large for particles with energy between $3 \times 10^{18}$  and
$6 \times 10^{19}$ eV, the calculations predict a large number of
extragalactic sources to contribute to the measured flux in this
energy range. This is a primary prediction of these calculations.

Of special interest is also the transition between galactic to
extragalactic predominance which should happen in the energy range
between $10^{16}$ to  about $10^{18}$ eV. Several experiments are
presently taking data in this energy range: KASCADE-Grande, Telescope
Array, IceTop and the Pierre Auger Observatory. The transition is a
very important feature because it is foreseen that breaks in the all
particle spectrum and in the composition can reveal the details of the
particle production mechanisms, the source population and propagation
in the Universe.  There are a number of recent attempts to explain the
cosmic ray spectrum in the transition range, e.g. by Hillas (Hillas
2006), and by Berezinsky et al. (Berezinsky et al. 2009).

In this paper, we focus on the previously inaccessible continuous
energy range above 10 PeV extending to the highest energies measured.
Today it is possible to compare the predictions with high precision
data over the entire energy range. Therefore it becomes important to
have predictive power, i.e. test quantitative hypotheses, which were
developed long before much of the new data was known.

We revisit here an idea originally proposed in 1993 (Biermann 1993;
Stanev et al. 1993) and we show how our Galaxy and the radio galaxy
Cen A can describe the energy spectrum from  10 PeV up to $3 \times
10^{20}$ eV and describe the galactic to extragalactic transition at
the same time.

In the following sections, we first go through the tests the 1993
original model has undergone to date as regards spectra, transport,
secondaries, and composition; secondly we confirm the predictions of
the original model with the newly available data beyond the knee
energy, and finally we present the high energy model which describe
the transition between Galactic and extragalactic cosmic rays.

\section{Original model and its tests to date}

In a series of papers started in 1993 (Biermann 1993; Stanev et
al. 1993; Biermann 1994) a astrophysics scenario was proposed which
emphasized the topology of the magnetic fields in the winds of
exploding massive stars (Parker 1958).  In (Stanev et al. 1993), a
comprehensive spectrum was predicted for six element groups
separately: H, He, CNO, Ne-S, Mn-Cl, Fe.  The key points of this
original model are:  a) The shock acceleration happens in a region,
which is highly unstable and shows substructure, detectable in radio
polarization observation of the shock region, also found in
theoretical exploration (e.g. (Bell \& Lucek 2001; Caprioli et
al. 2010; Bykov et al. 2011).  Therefore the particles go back and
forth across the shock gaining momentum, while the scattering on both
sides is dominated by the scale of these instabilities, which are
assumed to be given by the limit allowed by the conservation laws in
mass and momentum;    b) There are cosmic ray particles which get
accelerated by a shock in the interstellar medium, produced by the
explosion of a relatively modest high mass star, or, alternatively, by
a low mass supernova Ia. This  is most relevant for Hydrogen and less
for Helium and heavier nuclei;   c) Heavy cosmic ray nuclei derive
from very massive stars, which explode into stellar winds already
depleted in Hydrogen, and also in Helium for the most massive
stars. These explosions produce a two part spectrum with a bend that
is proposed to explain the knee. In this scenario the knee is due to
the finite containment of particles in the magnetic field of the
predecessor stellar wind, which runs as $\sin \theta /r$ in polar
coordinates (Parker 1958). Towards the pole region only lower energies
are possible and the knee energy itself is given by the space
available in the polar region.  There is a polar cap component of
cosmic rays associated to the polar radial field with a flatter
spectrum;    d) Diffusive leakage from the cosmic ray disk steepens
all these spectra by 1/3 for the observer; e) Very
massive stars eject most of their zero-age mass before they explode
and so form a very massive shell around their wind (Woosley
2002). This wind-shell is the site of most interaction for the heavy
nuclei component of cosmic rays. For stellar masses above about 25
solar masses in zero age main sequence mass (Biermann 1994) the
magnetic irregularity spectrum is excited by the cosmic ray particles
themselves. The spectral steepening due to the interactions is
$E^{-5/9}$ for the most massive star shells.

The final spectrum is a composition of these components, see Figure 1
of Ref. (Stanev et al. 1993). The spectra predicted by these arguments
match the data such as shown by the recent CREAM results (Wiebel-Sooth
et al. 1998; Biermann et al. 2009).  This scenario has undergone
detailed tests as regards propagation and interactions (Biermann 1994;
Biermann et al. 2009) so as to describe both Galactic propagation and
the spectra of the spallated isotopes as well as the resulting
positron spectra, the flatter cosmic ray positron and electron data,
the WMAP haze and the spectral behavior of its inverse Compton
emission, and the 511 keV emission from the Galactic Center
region. New Tracer results (Obermeier 2011) are also consistent in
terms of a) the low energy source spectrum, b)  the energy dependence
of interaction, c)  a finite residual path-length at higher energy,
and d) a general upturn in the individual element spectra.  The newest
Pamela results (Adriani 2011) are also consistent with the 1993
original model, in which Hydrogen was the only element to have a
strong ISM-SN cosmic ray component, and so has a steeper spectrum than
Helium.


\subsection{A test beyond the knee}

This original model was proposed to explain the particles observed
above $10^{9}$ eV per nuclear charge. Here we first test the original
model to the KASCADE data. The most  accurate  measurement of the
energy spectrum in the knee energy range has been done by the KASCADE
experiment (KASCADE-Grande Coll. 2010).  Figure 1 shows for the first
time the comparison of the original model to the measured data from
KASCADE. KASCADE reconstructs the spectrum using two hadronic
interaction programs (QGSJet and Sibyll) in the analysis procedure. In
the figure we show the data and the original model, and also include
the ratio of the difference between original model and data divided by
the experimental error. For the ratio shown we only use one of these
interaction codes, as an example we use QGSJet.  The figure shows good
agreement between data and the original model to within the errors of
the data.  This confirms that the original model in its last remaining
energy range, where it had not previously been tested for lack of good
data. This is the first key result of this paper.


\subsection{Transport and interaction test}

One question, which invariably comes up is how this model deals with
propagation and spallation.  For this line of reasoning it is
important to note two aspects:
a) Plasma simulations, the Solar wind
data (Mattheaus \& Zhou 1989), the interstellar medium data (Rickett
1977, Spangler \& Gwinn 1990, Goldstein et al. 1995) as well as radio
galaxy data (Biermann 1989) all are consistent with the interpretation
that in an ionized magnetic plasma in near equipartition we have an
approximate Kolmogorov spectrum (Kolmogorov 1941) running without
break from very large scales down to dissipation scale.
b) Very
massive stars eject most of the initial zero age main sequence mass in
their powerful winds, which then builds up a correspondingly massive
shell at the outer boundary of the wind.  It is this massive shell the
cosmic ray loaded supernova shock encounters and interacts with.

All those cosmic ray particles accelerated by a supernova shock in the
heavily enriched winds of Wolf-Rayet stars then excite magnetic
irregularities, which can be described following Bell (1978); it is
these self-excited irregularities that describe the path of the cosmic
ray particles through the massive shell.  This gives (Biermann 1998,
Biermann et al. 2001) a Boron/Carbon ratio energy dependence of
$E^{-5/9}$; this was found to be consistent with data by Ptuskin et
al. (Ptuskin et al. 1999), who determined $E^{-0.54 }$.  Since the
straight-line path gives a minimum path, this model also gives a
finite path-length of interaction at high energy, consistent with the
new data as noted above.

Other tests are: The cosmic ray electron spectrum has been determined
to be $E^{-3.23 \pm 0.06}$ (Wiebel-Sooth \& Biermann 1999) in the
energy range up to a few TeV, a spectrum which is dominated by losses
(Kardashev 1962).  Therefore the injection is with a spectrum flatter
by unity.  This has to be compared with the proton spectrum at a
corresponding energy, which suggests $E^{-2.66 \pm 0.02}$ near TeV
energies (Yoon et al. 2011); this spectrum has been steepened by the
energy dependence of the diffusion, and so the difference to the
inferred cosmic ray electron spectrum gives this dependence as
$E^{-0.43 \pm 0.06}$, consistent with $E^{-1/3}$ as deduced from the
Kolmogorov assumption.  Another consistency check is the time scale
inferred at the highest energies for leaking out of the kpc-thick
cosmic ray disk near the Sun (Biermann 1993).

Since we have about
$10^{7}$ yrs at GeV energies for protons, we infer with the Kolmogorov
assumption a time scale of $10^{4.3}$ yrs at $10^{17}$ eV - adopting
the point of view that the highest energies cosmic ray
particles from the Galaxy are Fe (Stanev et al. 1993) at $10^{18.5}$
eV, matching in their scattering protons of $10^{17}$ eV; however,
this is not yet finally settled.

This time
scale is still significantly longer than the simple transit time
across the thick disk of about $10^{3.5}$ yrs, so that the isotropy
observed can be understood without already invoking the effect of the
Galactic wind.  Using three times the simple transit time as the
minimum to give isotropy we can invert this line of reasoning and
deduce a maximum energy dependence of the scattering of $E^{-0.38}$
(Biermann 1993), under the assumption again, that the entire energy
range is covered by the same powerlaw.

The Wolf-Rayet star model is able to also explain the positron spectra
and positron production (Biermann et al. 2009, 2010) as noted above.

\section{The high energy model}

Based on the original model we propose here a high energy model to
explain the cosmic ray data from 10 PeV to 300 EeV.  We analyze the
possibility that the \emph{very same spectrum} proposed in 1993
however shifted in energy can explain all the data up to 300 EeV
including the knee region, the highest energy range ($> 10^{18}$ eV)
and at the same time the middle energy range ($10^{16} < E < 10^{18}$
eV).

The cosmic ray particles as seen in our galaxy were argued to provide
the seed particles for further acceleration to ultra-high energy by a
relativistic shock emanating from an active black hole in the nearby
radio galaxy Cen A (Gopal-Krishna et al. 2010). This idea is explored here
beyond what has been proposed in (Gopal-Krishna et
al. 2010) which demonstrated
that Cen A can  provide a sufficient flux for the highest
energy particles  by working out the energetic particle flux traversed
by a shock surface in the jet of the radio galaxy Cen A with the
one-step further acceleration in a relativistic shock as proposed by
(Achterberg et al. 2001). They have used the spectral  shape of the
original model (Stanev et al. 1993) to fit the Pierre
Auger data, however the fit of the measured spectrum was not
constrained by the low energy spectrum as proposed in the original
model (Stanev et al. 1993).

The energy spectrum calculated here is simply a shift of the particle
spectra proposed in the original model (Stanev et al. 1993) for low
energies to the highest energies preserving the relative abundances of
the  original model. This proposal is considerable stronger than
the previous one presented in (Gopal-Krishna et al. 2010).
The energy shift corresponds to a factor of 2800 within the limits of
a one-step acceleration by a relativistic shock as proposed by
(Achterberg et al. 2001).  The original
model (Stanev et al. 1993) has a number of parameters, which were set
in 1992, see Fig.6 in (Stanev et al.1993). None of these parameters
had to be changed significantly in the analysis presented here.

\section{Match to data from 10 PeV to 300 EeV}

Finally, we can construct the energy spectrum of cosmic rays by adding
the galactic component to the extragalactic component as shown in
figure 1. In this figure we show differential flux $\times E^{3}$  versus energy
per particle as predicted by this analysis compared to the data from
KASCADE (KASCADE Coll. 2009),  KASCADE-Grande
(KASCADE-Grande Coll. 2010) and the Pierre Auger Observatory
(Pierre Auger Coll. 2010a). We have shifted  within the experimental
uncertainties the KASCADE and
KASCADE-Grande flux down by 14\% and  the Auger flux up by 14 \% in
order to match.

We distinguish six groups, H, He, CNO, Ne-S, Cl-Mn, and
Fe.  Particles were subjected to the losses in the intergalactic
radiation field (Allard et al. 2008).  The numbers above the
lines correspond to error estimations (Model - Data)/(Experimental
Error). We note the good agreement between data and model from
below $10^{15}$ to $3 \times 10^{20}$ eV.

One extra assumption of the original model, namely the energy shift,
allows a description of the energy spectrum above $10^{18}$ eV.
Below the critical energy for interactions with the microwave
background, it has been expected that we would observe a very
large number of sources, at large distances. However, no other source
population is needed to describe the energy spectrum above $3 \times
10^{16}$ eV. This is the second key new result of this paper.


To summarize, a few results can be extracted from the proposal here:
I) There are no other sources necessary to provide extra flux in the
energy range $3 \times 10^{16}$ and $3 \times 10^{20}$ eV, the second
key result of this paper. A detailed analysis of radio galaxies
(Caramete et al. 2011)  shows that the next strongest radio galaxy
to contribute is Virgo A, as already predicted by Ginzburg \&
Syrovatskii (1963).  We estimate the maximum possible extra flux from
other sources within this distance limit at 0.1 in the log in Fig. 1,
so at 25 percent.  We note that the self-consistent MHD-simulations
for cosmological magnetic fields presented in (Ryu et al. 2008) were
carried out for protons and so, for heavy nuclei  the magnetic horizon
in intergalactic space is small, less than 100 Mpc consistent with the
measurements of the Pierre Auger Observatory (Pierre Auger
Coll. 2010c) which suggest a heavier composition for energies above
$10^{19}$ eV.  However, at yet lower energies the sum of the more
distant sources might exceed the flux predicted from single sources
such as Cen A and Vir A;  the magnetic scattering as predicted in (Ryu
et al. 2008) seems to prevent this at all energies above the
transition to Galactic cosmic rays;    II) The dip near $3\times
10^{18}$ eV is explained by the switch-over between the galactic
cosmic rays and the extragalactic cosmic rays (Rachen et al. 1993);
III)  The spectra of Galactic cosmic rays beyond the knee are
adequately modeled by our approach suggesting that the Wolf-Rayet star
explosion model matches also the newest data beyond the knee;   IV)
There is no abrupt change in composition in the energy range from $3
\times 10^{16}$  to $3 \times 10^{18}$ eV;

In order to describe the data (Pierre Auger Coll. 2010a) the high
energy model presented here requires minimal interaction along the
path between the radio galaxy Cen A and us, and so indeed near
isotropic scattering in the magnetic wind of our Galaxy.

\subsection{Isotropy?}

The most recent results from the
Pierre Auger Observatory indicate an excess in the direction of Cen
A (Pierre Auger Coll. 2010b) which might corroborate the analysis
presented here. The Pierre Auger Observatory measures an isotropic sky
for energies below $6 \times 10^{19}$ eV and a weakly anisotropic sky above
this energy (Pierre Auger Coll. 2010b). The isotropy for energies
below $6 \times 10^{19}$ eV can be explained by a turbulent magnetic
wind of our Galaxy (Everett et al. 2008). This wind is thought to be
driven by cosmic rays and hot gas, and so is itself unstable, giving
irregular magnetic fields, akin to radiation driven winds of stars,
(Cassinelli 1994; Owocki 1990).  The magnetic fields in the wind are strong
enough to isotropize incoming heavy element particles at very high
energy and protons at lower energy.  Scattering in a magnetic wind
with $B_{\phi} \sim 1/R$ as a function of radial distance $R$ (Parker
1958), is strongly
enhanced due to the extra factor derived from integrating the Lorentz
force  $\ln \{R_{max}/R_{min}\} \, \sim \, 5$.  The maximal energy for
total bending can then be given by the magnetic field strength
(Everett et al. 2008) at the base $\sim 8$ $\mu$Gauss, the length
scale at the base $\sim 5$ kpc, this logarithmic factor $\, \sim 5$,
and so is given by $10^{20.2} \, Z$ eV, where $Z$ is the charge of the
cosmic ray particle.  Since at those energies the data measured by the
Pierre Auger Observatory suggest
that we actually have heavy nuclei, complete bending is assured.
However, this would lead to a second problem, in that we then might
find a complete shielding for any particles coming from outside, so
this magnetic wind must also have considerable irregularities; these
irregularities in the wind need to be scale-free (implying a saturated
spectrum of irregularities, or inverse cascade $I(k) k \sim const$,
where $I(k)$ is the energy per wave number $k$ per volume), so as to
avoid a characteristic energy, below which all particles are cut off;
or, if such an energy exists, it must be low enough not to disturb the
spectral sum.  The key point is that Parker-winds (Parker 1958) are
very effective at bending orbits.  Obviously, the scattering might not
be complete, so that a small anisotropy is left possibly explaining
the Auger data clustering of events near the direction to the radio
galaxy Cen A.

\section{Predictions}

Some predictions of the high energy model presented here are:

1) The calculations presented here predict the individual spectra for
six element groups. The future data analysis from KASCADE-Grande,
IceTop (Stanev 2009), Telescope Array and the Pierre Auger Observatory
will be able to test this prediction;
2) The trend to heavier nuclei from 2 to $6 \times 10^{19}$ eV has
been suggested by measurements of the depth of shower maximum
done by the Pierre Auger Observatory (Pierre Auger Coll. 2010c); we
caution, that the interpretation of the data in terms of mass
composition depends on hadronic interaction extrapolations;
3) An isotropic background contribution of high energy cosmic rays
from other more distant sources is compatible with our analysis up to
25 percent of the total flux.


An important caveat of the analysis refers to gamma ray bursts. We
could obtain similar results in describing the cosmic ray spectrum if
instead of exploding Wolf-Rayet stars we would have used gamma ray
bursts exploding into Wolf-Rayet star winds.   This assumption allows
that Wolf-Rayet star explosions might be due to the same  mechanism as
gamma ray bursts, but just completely stifled by the mass burden,
possibly implying the magneto-rotational mechanism of
Bisnovatyi-Kogan (Bisnovatyi-Kogan 1970).  This would suggest that
most Galactic cosmic rays be attributed to gamma ray bursts (Dermer
2004), and that the gamma ray burst rate is substantially higher than
heretofore believed.   The
predictive power for the spectral indices and the energy scales is
lost in this alternative.

Finally, we show how the high energy model proposed here could be
falsified:
A)  If all ultra high energy cosmic rays could be shown to be of one
and only one chemical element, like all Proton, or all Iron;
B)  If neutrino or gamma ray data would unequivocally show that many
nearby extragalactic sources contribute equivalently to the radio
galaxy Cen A. This could occur naturally in a gamma ray burst
hypothesis, since many nearby starburst galaxies with high rates
of star formation, supernova explosions, and gamma ray bursts could
all contribute at comparable levels (Caramete et al. 2011);
C)  If it could be clearly shown that the turbulent magnetic wind of
our Galaxy does not have the required strength of magnetic field and
spatial extent to effect near isotropy by magnetic scattering;
D) If an abrupt change of the composition is measured between the iron
knee and the dip or ankle.

\section{Conclusions}

We conclude that our own Galactic cosmic rays plus the galactic cosmic
rays from a radio galaxy shifted in energy in the relativistic shock
of an accreting super-massive  black hole  reproduces the all particle
energy spectrum from $10^{15}$ to $10^{20}$ eV as measured by the
KASCADE, KASCADE-Grande and the Pierre Auger Observatories.  In the
scenario proposed here no additional extragalactic source population
for ultra high energy particles is required, contrary to many years of
expectation. That implies that no other sources within the magnetic
horizon is
viable even at lower particle energies ($> 3 \times 10^{18}$ eV)
above the switchover between galactic and extragalactic cosmic rays.

The scenario proposed here gives a number of predictions, especially
as regards the chemical element composition across this entire energy
range. An detailed comparison of the measured and predicted
composition is yet to be done. Once these predictions have been
falsified or confirmed we  will be closer to an understanding of the
origin of cosmic rays 100 years after their discovery.

\acknowledgments

PLB acknowledges discussions with J. Becker, J. Bl{\"u}mer,
L. Caramete, R.  Engel, J. Everett, H.  Falcke, T.K. Gaisser,
L.A. Gergely, A.  Haungs, S. Jiraskova, H.  Kang, K.-H.  Kampert,
A. Kogut, Gopal Krishna, R. Lovelace,  K.  Mannheim, I. Maris, G.
Medina-Tanco, A. Meli, B.  Nath, A.  Obermeier, J. Rachen,
M. Romanova,  D.  Ryu, E.-S.  Seo, T. Stanev, and P.  Wiita.  VdS and
PLB both acknowledge their KASCADE, KASCADE-Grande and Pierre Auger
Collaborators.  VdS is  supported by FAPESP (2008/04259-0,
2010/07359-6) and CNPq.


\clearpage

\begin{figure}
\includegraphics[scale=0.9]{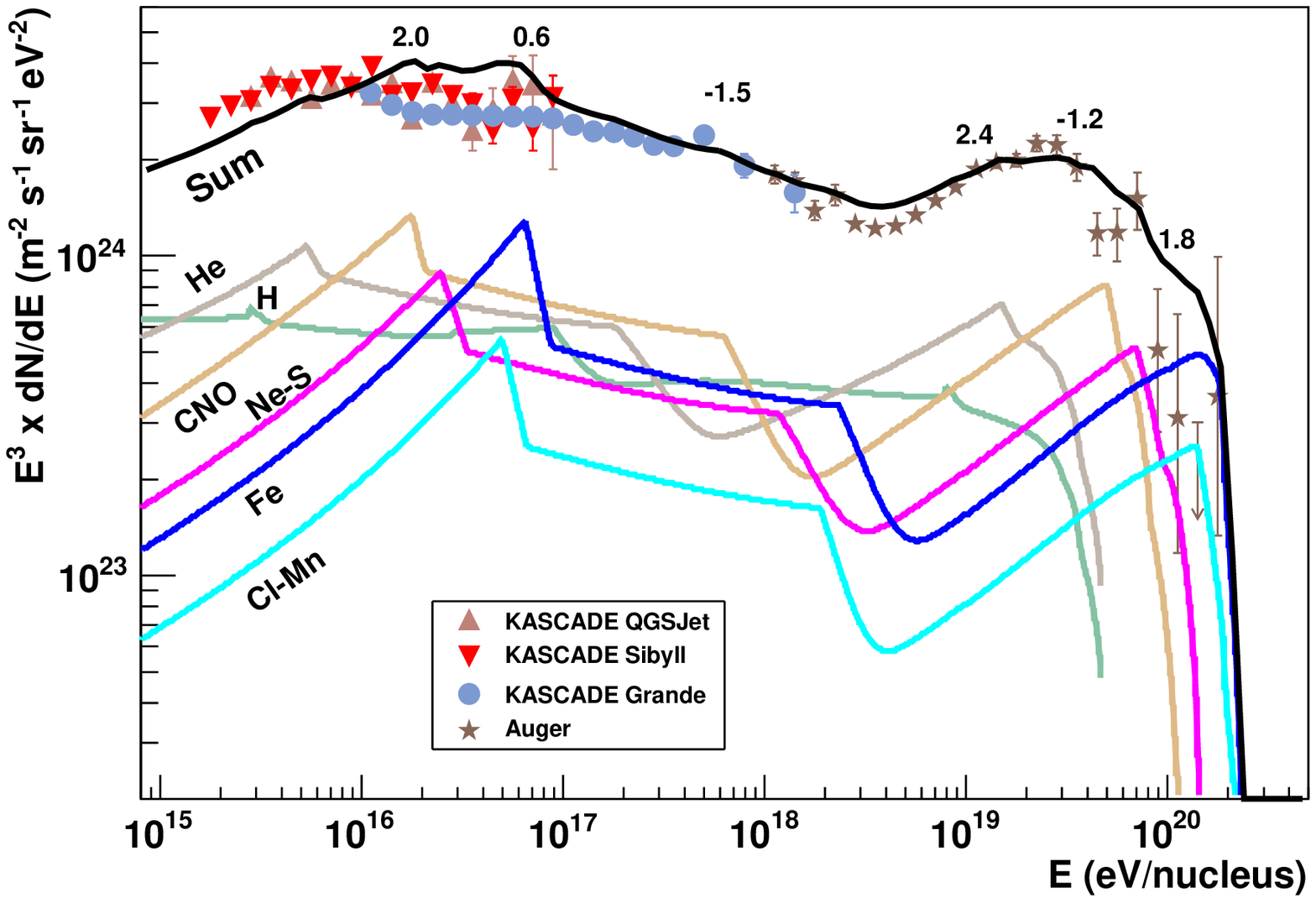}
\caption{The energy spectrum calculated with this model compared to
  the data from KASCADE (KASCADE Coll. 2009) ,
  KASCADE-Grande (KASCADE-Grande Coll. 2010) and Pierre Auger
  Observatory (Pierre Auger Coll. 2010a). The numbers in the upper part
  of the figure shows the error of the model defined as (Model -
  Data)/(Experimental Error). The shape of the six element spectra
  from the Galactic and the extragalactic component is the same, by
  model assumption.}
\end{figure}
\label{fig:spectrum:all}

\clearpage


\begin{thebibliography}{}


\bibitem{Achterbergetal01}  Achterberg, A., et al. 2001, MNRAS, 328, 393.


\bibitem{Pamela11} Adriani, O. (Pamela-Coll.) 2011, astro-ph:1103.4055.

\bibitem{Allard08} Allard, D., et al. 2008, \JCAP, 10, 33.

\bibitem{Anchordoquietal11} Anchordoqui, L.A., et al. 2011, astro-ph:1103.0536;
Fargion, D., \&  D'Armiento, D. 2011, astro-ph:1101.0273.

\bibitem{BaadeZwicky34} Baade, W., \& Zwicky, F. 1934,
  Proc. Nat. Acad.  Sci., 20, 259.

\bibitem{Bell78}  Bell, A. R. 1978a, MNRAS, 182, 147; MNRAS,  182, 443.

\bibitem{BellLucek01}  Bell, A. R., \& Lucek, S. G. 2001, \MNRAS, 321, 433.


\bibitem{Berezinsky}  Berezinsky, V. 2009, Nucl. Phys. B Proc. Suppl.,
  188, 227.

\bibitem{Biermannstrittmatter87} Biermann, P.L., \& Strittmatter,
  P.A. 1987, \ApJ, 322, 643.

\bibitem{Biermann89} Biermann, P. L., in Proc. {\it Hot spots in extragalactic radio sources}; Workshop, Tegernsee, 1989, {\it Lect. Not. Phys.} {\bf 327}, 261

\bibitem{Biermann93} Biermann, P.L. 1993, \AA, 271, 649;
  idem \& Cassinelli, J.P. 1993, \AA, 277, 691;
  idem \& Strom, R.G. 1993, \AA, 275, 659.

\bibitem{ICRC94} Biermann, P.L. 1994 in Proc. ``Invited, Rapporteur and
  Highlight papers", 23rd ICRC Calgary; Eds. D.  A. Leahy et al.,
  World Scientific, Singapore, p. 45.

\bibitem{Hirrschegg} Biermann, P.L. 1998, in Proc. {\it Nuclear Astrophysics}
meeting at Hirschegg, GSI, Darmstadt, p. 211.

\bibitem{Biermann2001} Biermann, P.L., et al. 2001, \AA, 369, 269.

\bibitem{Biermannetal09b} Biermann, P. L., et al. 2009 \PRL, 103,
  061101.

\bibitem{Biermannetal010a} Biermann, P. L., et al. 2010a, \ApJL, 710, L53.

\bibitem{Biermannetal010b} Biermann, P. L., et al. 2010b, \ApJ, 725, 184.


\bibitem{BK70} Bisnovatyi-Kogan, G. S., 1970, Astron. Zh., 47, 813;
  transl. 1971, Sov. Astron., 14, 652; Biermann, P.L., et al. 2005,
  AIP Proc., 784, 385; Bisnovatyi-Kogan, G. S., et al. 2005, astro-ph:0511173.


\bibitem{Bykov11} Bykov, A. M., et al. 2011 \MNRAS, 410, 39.

\bibitem{Blasi10}  Caprioli, D., et al. 2010, \ApP, 33, 307.

\bibitem{CarameteB11}  Caramete, L., et al. 2011, astro-ph:1106.5109.

\bibitem{Cassinelli}  Cassinelli, J. P. 1994, Astroph. \& Sp. Sci.,
  221,483.

\bibitem{DermerGRB} Dermer, Ch.D. 2004, in Proc. 13th Course of the
  Int. School of Cosmic Ray Astrop.; Eds: M.M. Shapiro, T. Stanev, \&
  J.P. Wefel, World Scientific, p. 189.

\bibitem{Drury83}  Drury, L. O'C. 1983, Rep. Progr. Phys., 46, 973.

\bibitem{Everettetal08}  Everett, J., et al. 2008, \ApJ, 674, 258.

\bibitem{Fermi49}  Fermi, E. 1949, \PR, 75, 1169;
 1954, \ApJ, 119, 1.

\bibitem{GaisserStanev08} Gaisser, T.K. \& Stanev, T. 2008, in Review of
  Particle Physics, \PLB, 667, 1.


\bibitem{Ginzburgsyrovatskii63} Ginzburg, V. L., \& Syrovatskii,
  S. I. 1963, Astron. Zh., 40, 466; transl. in 1963, Sov. Astron. A.J., 7, 357;
  1964, The origin of cosmic rays, Pergamon Press, Oxford, orig. Russ. ed. (1963).

\bibitem{Goldstein}  Goldstein, M. L., Roberts, D. A., Matthaeus,
  W. H. 1995, \ARAA, 33, 283.

\bibitem{GKetal10}  Gopal-Krishna, et al. 2010, \ApJL, 720, L155.

\bibitem{Greisen66}  Greisen, K. 1966, \PRL, 16, 748.

\bibitem{Hess12} Hess, V.F. 1912, Physik. Z., 13, 1084.


\bibitem{Hillas}  Hillas, A. M. 2006 astro-ph:0607109.

\bibitem{Kardashev62} Kardashev, N. S. 1962, Astron. Zh., 39, 393;
  transl. 1962, Sov. Astron. A.J., 6, 317.


\bibitem{bib:kascade:spectrum} KASCADE Coll. 2009, \ApP, 31, 86.

\bibitem{bib:kascade:grande:spectrum} KASCADE-Grande Coll. 2010
  astro-ph:1009.4716.

\bibitem{Kolmogorov41}  Kolmogorov, A. 1941, Dokl. Akad. Nauk SSSR,
  30, 299;  31, 538; and 32, 19.

\bibitem{kohlh} Kohlh{\"o}rster, W. 1913, Physik. Z., 14, 1153.

\bibitem{LagageCesarsky83} Lagage, P. O. \& Cesarsky, C. J. 1983, \AA,125, 249.


\bibitem{Lovelace76} Lovelace, R. V. E. 1976, \Nature, 262, 649.




\bibitem{Matthaeus89} Matthaeus, W. H. \& Zhou, Y. 1989, Physics of
  Fluids B, 1,1929.

\bibitem{Tracer11}  Obermeier, A. 2011, Ph.D. thesis, Radboud University
  Nij\-megen.



\bibitem{Owocki} Owocki, S. P. 1990,Rev. Mod. Astron., 3,98.

\bibitem{Parker58}  Parker, E.N. 1958, \ApJ, 128, 664.



\bibitem{bib:auger:spectrum} Pierre Auger Coll. 2010a,
  Phys. Lett. B, 685, 239.

\bibitem{bib:auger:anisotropy} Pierre Auger Coll. 2010b, \ApP,34, 314.


\bibitem{bib:auger:xmax}  Pierre Auger Coll. 2010c, \PRL 104, 091101.

\bibitem{}  Ptuskin, V., Lukasiak, A., Jones, F.C., Webber, W.R. 1999, ICRC Salt Lake City, vol. 4, p. 291

\bibitem{Rachen93}   Rachen, J.P., et al. 1993, \AA, 272, 161;
 1993, \AA, 273, 377.

\bibitem{Rickett77}  Rickett, B.J. 1977, \ARAA, $\;$ 15, 479.

\bibitem{Ryuetal08}  Ryu, D., et al., 2008, Science, 320, 909);
Das, S., et al. 2008, \ApJ, 682, 29;
Cho, J., \& Ryu, D. 2009, \ApJL, 705, L90.

\bibitem{Spangler} Spangler, St. R., \& Gwinn, Carl R. 1990, \ApJL, 353, L29.

\bibitem{Stanevetal93} Stanev, T., et al. 1993, \AA, 274, 902.

\bibitem{bib:icetop} Stanev, T. (IceCube Coll.) 2009
  astro-ph:0903.0576.

\bibitem{Stanev10a}  Stanev, T. 2010a Review at Vulcano Workshop
  2010, astro-ph:1011.1872;

\bibitem{Stanev10b} Stanev, T. 2010b, High Energy Cosmic Rays, Springer.

\bibitem{Vietri95} Vietri, M. 1995, \ApJ, 453, 883.

\bibitem{VoelkBiermann88} V{\"o}lk, H.J. \& Biermann, P.L. 1988, \ApJL, 333, L65.


\bibitem{Waxman95} Waxman, E. 1995, \PRL, 75, 386.

\bibitem{WSBM98}  Wiebel-Sooth, B., et al. 1998, \AA, 330, 389.

\bibitem{WSB99}  {\it Cosmic Rays}, Wiebel-Sooth, B., \& Biermann, P.L.,
   in Landolt-B{\"o}rnstein, Handbook of Physics, Springer
  Publ. Comp.,  p. 37 - 91, 1999

\bibitem{Woosley02}  Woosley, S. E., Heger, A., Weaver, T. A. 2002,
  \RMP, 74, 1015.

\bibitem{Yoon2011}  Yoon, Y.S., et al. 2011, \ApJ  {\bf  728}, id.122

\bibitem{ZatsepinKuzmin66}  Zatsepin, G. T., Kuz'min, V. A. 1966,
  Zh. Exp. Th. Fis. Pis'ma, 4, 114; engl. transl. 1966  J. of Exp. \&
  Th. Phys. Lett., 4, 78.



\end{thebibliography}
\end{document}